\documentclass[12pt,a4paper]{article}
\usepackage{amsfonts,amssymb,amsmath,graphicx} \usepackage{bm} \usepackage[spanish,english]{babel}
\usepackage[latin1]{inputenc}
\usepackage{hyperref}

\begin{document}

\def\bce{\begin{center}} \def\ece{\end{center}} \def\beq{\begin{eqnarray}} \def\eeq{\end{eqnarray}}
\def\ben{\begin{enumerate}} \def\een{\end{enumerate}} \def\bei{\begin{itemize}}
\def\eei{\end{itemize}} \def\ul{\underline} \def\ni{\noindent} \def\nn{\nonumber \\}
\def\bs{\bigskip} \def\ms{\medskip} \def\tr{\mbox{tr}} \def\wt{\widetilde} \def\wh{\widehat}
\def\brr{\begin{array}} \def\err{\end{array}} \def\dsp{\displaystyle} \def\eg{{\it e.g.}}
\def\ie{{\it i.e.}}

\title{Reasons in favor of a Hubble-Lemaître-Slipher's (HLS) law}

\author{{\large Emilio Elizalde} \\ {\small National Higher Research Council of Spain}
\\ {\small Institute of Space Sciences (ICE/CSIC and IEEC)} \\ {\small Campus UAB, Carrer de
Can Magrans, s/n, 08193 Bellaterra (Barcelona), Spain}  \\ {\small e-mail: elizalde@ieec.uab.es} \\ }

\maketitle

\begin{abstract}

Based on historical facts, revisited from a present-day perspective, and on the  documented opinions of the scientists involved in the discovery themselves, strong arguments are given in favor of a proposal to include prominent astronomer Vesto Slipher to the suggested addition of Georges Lemaître's name to Hubble's law on the expansion of the Universe, and thus eventually call it Hubble-Lemaître-Slipher's (HLS) law.

\end{abstract}



\section{Introduction}

To find the origins of Astronomy and even of Cosmology itself, that is, of the knowledge of the Universe on a large scale, as a whole, we would have to go far back in the history of humanity. It is generally accepted, however, that modern cosmology, as a science, began to take shape at the beginning of the last century. Before that, the first modern scientists appeared, being Galileo Galilei (1564-1642), according to general consensus, the first of them. Much we could talk about Galileo, but this is not the priority objective of this article, so I will limit myself to refer only to a couple of aspects that do have to do with him. Galileo taught us that Science must be based on two fundamental pillars, namely, the experimental observation of nature and the scientific theory itself. If either of them fails, there is no Science. However beautiful a scientific {\it theory} may be, by ``natural'' or ``evident'' that appears to our eyes in a thousand aspects, if it is not endorsed by observation of nature, with precise experimentation, it will be wholly worthless.
At the other extreme, a table of results of our observation of nature, however accurate they may be, and however complete the table, will remain only in that: an empirical observation that will have a very limited value, perhaps phenomenological, very local and particularized. To promote these results at a much broader and more general level, the formulation of a law, of a theory, at least of a model, as generic as possible, is absolutely necessary.

That is precisely what Galileo did with his many experiments and his formulations of general principles of nature. He discovered very important laws of physics, among them the principle of relativity, which precedes the final formulation by Albert Einstein centuries later.
The scientific method requires a theory --- better as simple, beautiful and general, unique, if possible, within the field in question --- and of its experimental verification in the laboratory, as accurately as feasible and by different groups, totally independent of each other. As an example, the Theory of Relativity has more than satisfied, in a very bright, unequaled way, all these criteria. Starting from pure logic, from basic assumptions that cannot be demonstrated, but apparently very clear and that Einstein took as axioms, as the equivalence of the inertial mass and the gravitational mass, and the constancy of the speed of light in vacuum --- proven in the experiment conducted in 1887 by Albert A. Michelson and Edward W. Morley --- and relying on the geometry of Bernhard Riemann and others, Einstein built a beautiful theory that has been proven true in all the experiments carried out to this day in our Solar System, in our galaxy, and in even more distant regions of our universe.

They were certainly developments of geometry in the XIX Century, in particular of the non-Euclidean geometries, on the part of Friedrich Gauss (1777-1855), Nikolai Lobachevski (1792-1856), János Bolyai (1802-1860) and Bernhard Riemann (1826-1866), which allowed Einstein --- with the inestimable collaboration, according to some historians, of his mathematical partner Marcel Grossmann and his wife Mileva Maric (1875-1948) --- to formulate his theories of Relativity, first the Special and then the General one. We must not forget the important contributions of other great scientists, such as Henri Poincaré (1854-1912), David Hilbert (1862-1943) and Hendrik Lorentz (1853-1928), to mention only three of them. In short, all modern cosmology is based on Einstein's field equations, which he formulated in 1915, corresponding to his theory of General Relativity (for interesting recent overviews, see e.g., \cite{ior15, deb16}). This gives rise to many authors to place, with precision, the origin of modern cosmology in 1917, in which Einstein used his equations, for the first time, to build a model with which to describe our Universe. The universe, at that epoch, was believed to be static, without origin or end in time, and was very small because everyone was convinced that everything discovered was within the Milky Way. Since a static universe was not a solution to his field equations, Einstein introduced into them an additional term, the now famous cosmological constant, which with the right sign provided a kind of repulsive force that counteracted the attraction of gravity.

Reflecting now for a moment, however little we know about physics we will be able to understand that a universe in uniform expansion, like the Big Bang model, does not need the action of any force to continue like this, expanding indefinitely, being sufficient for it the initial impulse of the Big Bang. This will happen, of course, as long as the matter/energy density of the universe is lower than a certain critical value; since, otherwise, a high density of it would be able to slow down the expansion of the universe and to reduce it to zero, at which point it would begin to contract until it would end in what is known as the Big Crunch. Being even more pedagogical, imagine a child, say the Little Prince, who lives on a minuscule planet of the size of a small mountain. If the Little Prince threw a stone with all his strength it would not return and then fall on him, but would be lost in space, moving away forever. But if the Little Prince would throw it from the Earth, it is sure that the stone would end up falling. It is the same physical principle that I am talking about: gravitational attraction. Once again: a Universe with a matter density lower than a critical value would never be able to stop its expansion, which would continue forever (to find this critical value and how much the actual mass/energy density of our Universe deviated from it was the main question in cosmology for decades).

Actually, in 1922 Alexander Friedmann found a solution of Einstein's equations, which corresponded to an expanding universe. He wrote to Einstein telling him about his remarkable finding, but Einstein did not buy such possibility. For years he opposed this crucial idea, which is the key issue of this essay: the history of the discovery of the expansion law of our Universe.


Very recently, at the	XXXth General	Assembly of the International Astronomical Union (IAU),	celebrated in Vienna (August 20th to 31st,	2018), five	Resolutions	were	proposed for approval \cite{IAU1}. The Fifth of them, Resolution B4, addressed a suggested renaming of the Hubble Law, recommending that from now this law on the expansion of the universe be referred to as the ``Hubble-Lemaître law''. The basic point in favor of the resolution being \begin{quotation}
that the Belgian astronomer Georges Lemaître, in 1927 published (in French) the paper entitled {\it `Un Univers homogene de masse constante et de rayon croissant rendant compte de la vitesse radiale des nébuleuses extra-galactiques'} \cite{lem1}. In this, he first rediscov­ers Friedmann's dynamic solution to Einstein's General Relativity equations that describes an expanding universe. He also derives that the expansion of the universe implies the spectra of distant galaxies are redshifted by an amount proportional to their distance. Finally he uses published data on the velocities and photometric distances of galaxies to derive the rate of expansion of the universe (assuming the linear relation he had found on theoretical grounds)''.
\end{quotation}

In the 	supporting	 Bibliography, for	convenience	of	the	 voters, 	an	excerpt	from	the	paper	by	David L. Block, {\it ``Georges	Lemaître and	Stigler's Law of Eponymy''}, reporting on	 an	interesting 	comment	on	the	matter	by	Lemaître	himself, was highlighted: ``In	a	Comment	published	in	Nature	\cite{mliv1}	Mario	Livio	has unearthed	a	letter	from	Lemaître	to	W.M.	Smart	(dated	9	March	1931).	From	that document,	it	is	clear	that	Lemaître	himself	translated	his	1927	paper	into	English	and who	also	omitted	his	determination	of	the 	coefficient	of	expansion	of	the	Universe ($H_0$) from values	of	radial	velocities	available	as	of	1927.	However, in	his	Comment
Livio	omits	a	vital	reference,	namely	thoughts	penned	by	Lemaître	himself	in	1950 \cite{lem2}:
\begin{quote}
About	my	contribution	of	1927,	I	do	not 	want	to	discuss	 if	I	was	a	professional	astronomer. I	was,	in	any	event,	an	IAU	member	(Cambridge,	1925),	and	I	had	studied	astronomy	for	two years,	a	year	with	Eddington	and	another	year	in	the	 U.S.	observatories. I	visited	Slipher	 and Hubble	and	heard	him	in	Washington,	in	1925,	making	his	memorable	communication	about the	distance	[to]	the 	Andromeda	nebula.	While	my	Mathematics	bibliography	was	seriously in	default 	since	I	did	not 	know	the	work	of	Friedmann,	it	is	perfectly	up	to	date	from	the astronomical	point	of	view;	I	calculate	[in	my	contribution]	the	coefficient	of	expansion (575	km	per	sec	per	megaparsec,	625	with	a	questionable	statistical	correction).	Of	course, before 	the	 discovery	and	study	of	clusters	of	nebulae,	there	was	no	point	to	establish the Hubble 	law,	but	only	to	calculate	its	coefficient.	The	title	of	my	note	leaves	no	doubt	on	my intentions:	A	Universe	with	a	constant	mass	and	increasing	radius	as	an	explanation	of	the radial	velocity	of	extra-galactic	nebulae.	
\end{quote}
In	1950,	Lemaître	clearly	did	not	want	the	rich	fusion	of	theory	and	observations contained	in	his	1927	paper	to	be	buried	in	the	sands	of	time.''

The	discussion	on	the 	Resolution	  B4	was	very	lively	but it	had	to	be	stopped		in	order	to	keep up with the	schedule, in particular, the subsequent	Closing	Ceremony. Some additional questions	were sent	by	email, as this one: \\

Q. Should	other	contributors	to	the	data used	in	the	early	expansion law  (Slipher, Leavitt, Stromgren, ...) be acknowledged as well? \\

A. No	because	they	did	 not 	use 	their	data	nor	invent	new	theory	to	discover	the 	Universal Expansion. \\

The author is essentially in agreement with all the considerations above, as they were formulated; in particular, with the last one, which refers to Slipher and Leavitt. Sure, one cannot object the sentence, as formulated, that these prominent astronomers ``did not use their data'' (in particular, Slipher, even if he calculated practically all redshifts used subsequently both by Hubble and Lemaître) or methods (Leavitt's law) profusely used  by Hubble as his main tool to obtain all  of his distances, ``in order to invent a 	new	theory	to	discover	the	 Universal Expansion.'' However, some crucial historical facts, which have been largely overseen till now, and which the author has rescued from oblivion, containing  documented opinions of the scientists involved (Hubble, in particular, on several occasion), as well as more recent bibliographical studies (duly mentioned below), have led him to formulate strong arguments in favor of a claim to include the prominent astronomer Vesto Slipher to the planned addition of Georges Lemaître's name to Hubble's law, to eventually call it Hubble-Lemaître-Slipher's (HLS) law.

 \section{Vesto Melvin Slipher}

It is quite easy to become convinced of how extremely difficult it is to calculate distances in astronomy. The order of magnitude of the mistakes committed in this respect during past epochs is overwhelming. Let us just recall that, in the first models of the cosmos (Anaximander's one, for instance) stars were considered to be closer to the Earth than the Sun \cite{anax1} (Sun was fire, and fire goes up; and the Sun was a bigger fire than the Moon or the stars). This is why the discovery done by Henrietta Swan Leavitt in 1912, after several years of collecting thousands of data, in particular from the Magellanic clouds, was so extraordinarily important. Namely the period-luminosity relationship of Cepheid variable stars: a linear dependence of the luminosity vs. the logarithm of the period of variability of the star's luminosity \cite{leavi1}. It would be interesting to describe the favorite physical mechanisms available to explain such relationship (as the Eddington valve \cite{edval1}, for a very beautiful one), but regretfully, there is no place here for that. Henrietta was a distinguished member of the so-called Edward Pickering's Harvard harem, better known as Harvard computers, a group of young ladies that did a tremendous job in astronomy at that time (interested readers can find more details in, e.g., \cite{leavi2}). Leavitt's result was an extremely powerful tool to calculate distances, in fact the main one employed by Hubble in the years to follow. And by several generations of astronomers, later, with enormous success, until other improved techniques appeared \cite{rr1}, most recently, the SNIa standardizable candles \cite{stcan1}, which led to the discovery of the acceleration of the Universe's expansion (Physics NP 2011). 

Because of the extraordinary difficulty to calculate distances, until 1923 the most extended belief was that the Milky Way contained the entire universe, which, on the other hand, had always existed and was static (for this is the natural final state of any physical system, under very general conditions and enough time to evolve).  The discovery of a Cepheid star in Andromeda \cite{cefpl1} allowed Hubble to determine, by using Leavitt's law, that this nebula was clearly outside of the Milky Way, and thus confirm the so-called Island Universe hypothesis. The conjecture of the possible existence of Island Universes had by then a long history already, going back at least to the 18th Century (see, e.g., M.J. Way \cite{way1}), with contributions by, among others, Swedenborg (1734), Wright (1750), Kant (1755), and Lambert (1761) \cite{lam1}. William Herschel (1785)  \cite{her1} was also convinced for a time that spiral nebulae were outside the Milky Way, but changed his mind later \cite{lund1}. 

In the very same year of 1912 that Leavitt had published her results, Vesto Slipher started a project aimed at obtaining the radial velocities of spiral nebulae from their spectral blue- or red-shifts, by using the 24-inch telescope of the Lowell Observatory, in Arizona. Actually, his very first calculation, which he produced on 17 September 1912, was for the Andromeda nebula, a blueshift \cite{sliph1}. In 1914, in a meeting of the American Astronomical Society, he presented results for a total of 15 nebulae. He was so convincing  that his results were received by the audience (chronicles say) with a very long, standing ovation \cite{sliph2}. This was really unusual, in a scientific conference, then as now, and that day went into the History of Astronomy \cite{sliph2a, Bremond2008, put1}. 

Slipher was the first to photographically detect galaxy spectra with
sufficient signal to noise ratio to reliably measure their Doppler shifts. 
As Hubble himself later recognized, he was the first astronomer to note that something highly remarkable, very strange, was going on in the cosmos: {\it how could the Universe be static with those distant nebulae receding at such enormous speeds?} Of course peculiar velocities were at play also, the dipole effect had to be taken into account, and later the translation speed of our own galaxy could be measured. But, already in these very first results, the general scattering trend of the far distant objects was apparent. It is thus clear why the potential importance of Slipher's discovery was so quickly appreciated by the attendees  \cite{sliph2a, Bremond2008, put1, jpea1}. 

In a previous paper of 1913 Slipher had measured the blueshift of Andromeda to be 300 km s$^{-1}$. In another, in 1914, he presented what seems to be the first demonstration that spiral galaxies rotate. An in his most famous one, of 1915, corresponding to the just mentioned AMS meeting of 1914, he reported on 15 nebulae, 11 of which were clearly redshifted, and the other four (the closest ones) blueshifted. By 1917 Slipher had 25 results, four of them blueshifts, and he provided an interpretation on the enormous receding mean velocity, of nearly 500 km s$^{-1}$, of these objects. He said namely that
\begin{quote} ``This might suggest that the spiral nebulae are scattering but their distribution on the sky is not in accord with this since they are inclined to cluster.''\end{quote} The term `scattering' already denotes a tendency to recede in all directions, which might led to the idea of an expanding universe \cite{jpea1} (see also \cite{jrgot1}). He added that:  \begin{quote} ``... our whole stellar system moves and carries us with it. It has for a long time been suggested that the spiral nebulae are stellar systems seen at great distances ... This theory, it seems to me, gains favor in the present observations.''\end{quote} In other words, Slipher rightly inferred that our galaxy was in motion at a very high speed, and that, most probably, the receding nebulae could be analogues of the Milky Way. And this was written eighth years before Hubble's detection of the famous Cepheid in Andromeda \cite{cefpl1}, which finally confirmed the `island universe' hypothesis.

Actually, Slipher did not repeat the same  analysis with the new table of redshifts, which were made public in 1923 on page 162, Chapter 5, of Arthur Eddington's book {\it The Mathematical Theory of Relativity} \cite{eddi1}, which was to become very popular.  In fact, Eddington had made a special effort to include a complete list of Slipher's redshifts in his book. He obtained them from Slipher through direct correspondence and consisted of 41 velocities, with only 5 blueshifts. By then, other astronomers had confirmed a number of Slipher's measurements, but it seems that the importance and physical meaning of a distance-redshift correspondence was not clear at all in the early 1920s. In 1924, Eddington  discussed the distribution of velocities and tried to make sense of them in the context of General Relativity, but he was still far from achieving what Lemaître did three years later \cite{lem1}. Actually, what Eddington believed was that  some kinematic effect was at work there, even if the most usual interpretation at the time involved the de Sitter model and was that the redshifts probed the curvature of spacetime. Nobody was thinking about the possibility of an expanding universe when looking for a relation between redshifts and distances. The aim was to search for the `de Sitter effect' and thus `measure the radius of curvature of spacetime' (see e.g., the 1924 references \cite{swl1}). There is a little exception to this since, in 1923, Hermann Weyl did conclude that ``space objects have a natural tendency to scatter'' \cite{weyl1}, but the reason for that `natural tendency' was unknown to him. And, even in 1929, Hubble also mentioned ``the de Sitter effect'' as well as Eddington's argument for a kinematical contribution, without ever saying that expansion could play any role \cite{hubb1}.

From what has been explained already, it is clear that Vesto Slipher was one of the pioneers of modern cosmology, and his contribution needs to be recognized as such. Also Edwin Hubble was a pioneer, but of a different kind, since in his case his obssession was to make sure that his legacy would prevail. He was very selective in referencing, failing to mention in his publications those of his colleagues (as is  known, the famous astronomer Harlow Shapley, one of the contenders of the `Great Debate' on Apr. 26, 1920 \cite{mbart1}, had a long complaint against Hubble on these matters).
But this is also precisely why the few, albeit very positive, comments Hubble actually made on the importance of Slipher's work (see below) are even much more valuable. Here it should be added that the importance of the redshift measurements and their significance in contradicting the static Universe model were also understood, by 1917, by other astronomers working at different observatories, not just by Slipher (at Lowell), but also by James E. Keeler (at Lick and Allegheny) and by William W. Campbell (at Lick). It was not an exclusive Mount Wilson result, as argued by Hubble for years.

Slipher's table of redshifts in Eddington's book was one of the two ingredients involved in the formulation of the distance-radial velocity relation. The other one, the table of distances, was the result of the work by Edwin Hubble, with a subsequent contribution of Humason, who obtained some additional redshifts, used by Hubble to improve his first results. 
Although Milton Humason extended the galaxy redshift work to fainter galaxies on behalf of Edwin Hubble, the astronomers at Mount Wilson would not have made rapid progress without Slipher's pioneering results. Hubble was fully aware of the significance and priority of Slipher's early spectroscopy \cite{put1}, but consistent with his style of claiming sole credit for most topics he worked on, he did not like to emphasize this point, only recognized later in his life (see next section).

William Hoyt  \cite{hoyt1980}, in his biographical memoir of V.M. Slipher (p. 411) claims that he \begin{quote} ``probably made more fundamental discoveries than any other observational astronomer of the twentieth century.''\end{quote} and John Peackok adds to that \cite{jpea1} \begin{quote} ``Slipher was indeed a great pioneer; not simply through his instrumental virtuosity in achieving reliable velocities where others had failed, but through the clarity of reasoning he applied. Slipher in 1917 lacked the theoretical prior of a predicted linear distance-redshift relation, which de Sitter only published the same year. Slipher was simply looking for a message that emerged directly from the data, and it is therefore all the more impressive that he was able to reach in 1917 his beautiful conclusions concerning the motion of the Milky Way and the nature of spiral nebulae as similar stellar systems. Slipher's other main legacy to modern cosmology remains as relevant as ever. The peculiar velocity field that he discovered has become one of the centerpieces of modern efforts to measure the nature of gravity on cosmological scales.''\end{quote} 


\section{Hubble on Slipher}

In his 1929 paper \cite{hubb1}, Hubble used a sample of red and blue-shifts of 24 nebulae, measured by Slipher; 20 of those were redshifts, with a maximum  of 1100 km s$^{-1}$. This was precisely the sample made available by Slipher in 1917. By adding to Slipher's velocities his measured distances, Hubble concluded that the mean velocity of the sample was positive (e.g., a recession, a redshift) and that a linear correlation between redshifts and distances was apparent. As Peackok explains in much detail \cite{jpea1}, actually ``Hubble was fortunate in a number of ways to have been able to
make such a claim with the material to hand. Hubble admitted that he was
following up previous searches for a distance-redshift correlation, and that these studies were explicitly motivated by the theoretical prior of the de Sitter effect.'' No universal expansion was in sight there.

Two years later, in 1931, Hubble and Humason pushed the maximum velocity by almost twenty, up to 20,000 km s$^{-1}$ \cite{huhu1}. However, their distance measurements were again based on Lundmark's  (unjustified) assumption of 1924, namely, that galaxies could be treated as standard objects for calculating their distances ---in the absence of other well-justified distance estimates. Again following Peacock, ``one could imagine that the 1931 paper should have received a good deal of critical skepticism, but by this time a linear D(z) relation was already regarded as having been proved.  The farther away a galaxy is, the faster it is moving. These results flew in the face of both Isaac Newton and Albert Einstein's notions of the universe, which argued for a static Universe. If Hubble was right, the visible objects of the universe were actually in expansion.'' 

This was certainly a very remarkable achievement, even if it was only a first step towards the modern conception: it did say nothing about the expansion of the Universe itself, of the fabric of spacetime. All what was said was just, that distant nebulae were receding from us with a velocity proportional to their radial distance, a totally empirical conclusion (see, please, the introduction again): the visible Universe was not static, it was expanding since distant objects were receding at very large speed. The `subtle' (but essential!) difference between two possible interpretations, namely (i) the celestial objects are moving away (cf. a child running away from us), or (ii) the objects have some small peculiar motions but are embedded in a fast moving reference frame (the child being now on a departing high-speed train and just maybe walking the corridor, to stretch his legs), what actually corresponds to an expanding solution of the universe model (Friedmann's solution of Einstein's General Relativity, in our case) was, and still is nowadays, for the non-specialist, a most crucial point (and not so easy to understand). For one, it took Einstein ten years to fully comprehend this concept in its final version, namely that the second interpretation was the right one, the one that is universally accepted nowadays.

Just to repeat, Hubble only contributed  with his own work, half of his famous plot, i.e. the distance measurements, and it so happens that his values turned out to be off by a whole order of magnitude, because of the incorrect distance measurements  of the stars used as standard candles. By contrast, the recession velocities on the y-axis of the plot, obtained by Vesto Slipher at the Lowell Observatory in Flagstaff, Arizona, had much smaller errors. Hubble used them  with permission from Slipher (they were already public, actually, as explained above), but gave no credit to him in the references of his 1929 work \cite{hubb1}. It is true that in the 1931 paper by Hubble and Humason \cite{huhu1} a generous acknowledgment to Slipher's contribution appears. However, in these two years that elapsed between the two papers, Hubble's name had been already associated with the distance-redshift correlation, and the fact without possible discussion is that Slipher's contribution was largely forgotten subsequently. And it is still being forgotten today, since Edwin Hubble continues very  often to be incorrectly credited (in many reference books, articles, talks and conferences, even by Nobel Prize awardees) with having obtained {\it both} the redshifts and the distances of the galaxies in his famous plot \cite{ee1}.

Later, however, having already acquired all his fame and by looking in retrospect, Hubble himself did recognize the extremely important role of Vesto Slipher, talking of \begin{quote} {\it ``your velocities and my distances''},\end{quote} in a Letter of E.P. Hubble to V.M. Slipher, Mar 6, 1953 [Biographical Memoirs, Vol 52, National Academy of Sciences (U.S.)]. This actually ocurred, as Marcia Bartusiak explains \cite{mbart2, mbart1}, as follows:  ``In 1953, as Hubble was preparing a talk, he wrote Slipher asking for some slides of his first 1912 spectrum of the Andromeda Nebula, and in this letter he, at last, gave the Lowell Observatory astronomer due credit for his initial breakthrough, writing: \begin{quote} {\it `I regard such first steps as by far the most important of all. Once the field is opened, others can follow.'}\end{quote} Also during the lecture, Hubble recognized that his discovery \begin{quote} {\it `emerged from a combination of radial velocities measured by Slipher at Flagstaff with distances derived at Mount Wilson.'}\end{quote}

Further, in his famous book \cite{hub_b1}, he repeated almost the same sentence, while referring to Slipher, saying once more that \begin{quote} {\it ``... the first steps in a new field are the most difficult and the most significant. Once the barrier is forced further development is relatively simple.''}\end{quote} All those quotes are explicit recognition by Hubble of the seminal role of Vesto Slipher in the derivation of the expansion law. Regretfully, this just happened too late (some of the quotes being from the  year when Hubble passed away), as we can witness nowadays (please re-read the sentence two paragraphs above).

Hubble was quite right in finally acknowledging the importance of Slipher's contribution. Although Slipher did not calculate the distances to the far away objects, his results on the redshifts, which he started to obtain in 1912, where very remarkable, as sufficiently emphasized above, pointing in some cases to very high recession speeds. He was undoubtedly the first astronomer to recognize that something weird, incredibly unusual was happening in the Universe: how could it be static at all? 
A very detailed account of Slipher's contribution can be found in \cite{orai1}.

\section{The expanding universe}

However, as already remarked, to say that the Universe is expanding because the far distant nebulae are receding at incredible speeds (this last sentence may be acceptable to everybody) is conceptually very far from the actual comprehension of the universe expansion as we understand it today: the stretching of the fabric of space and time itself, of the coordinate system of the cosmos. As conveniently emphasized in the introduction, in order to obtain a fundamental, physical explanation one needs a theoretical model to match the observational results; and it turned out that the de Sitter model, considered at first, was not the right one. 

That was not an easy challenge.
To the best minds of the time (as Eddington or Einstein), even when confronted with the astronomic observations and the theoretical model (see, e.g. \cite{ee1}, for a detailed account), it took several years to understand that Friedmann's expanding solution was the correct one. This may sound nowadays almost incredible, because Einstein was the father of the theory (GR) from which all the solutions emanated, and he had spent so much time and penetrated so deeply into the nature of the physical laws, and gravity in particular, to construct his magnificent theory of space and time. 

Actually it was no other than Georges Lemaître the first to understand what could be going on, in his now famous but for many years neglected paper of 1927 \cite{lem1}. He found a value for the slope of the linear relation between distance and velocity, and this {\it in the context of an expanding universe} (his re-discovered Friedmann's solution). And that happened two full years before Hubble published his famous law. Lemaître's  role in this story ---as the first who actually obtained Hubble's expansion rate and who did interpret this expansion as a true stretching of space--- has been already explained in detail in the Introduction (for more information, see also \cite{ee1}). 

From a more general and wider perspective, we shoud add that several other scientists, as Carl Wirtz, Ludwik Silberstein, Knut Lundmark, or Willem de Sitter himself, had been actually looking for a redshift-distance relation of a similar kind, which could fit into the context of de Sitter's model \cite{swl1} (see also \cite{way1}). As reported in \cite{bel1}, the first theoretical explanation capable of accounting for the 
Slipher's redshifts as Doppler effect was suggested by Alexander Friedmann in 1922 \cite{fri22} \begin{quotation} {\it 
``the Universe may expand since General Relativity equations admit dynamical solutions.'' }
\end{quotation}  Having been told of  this conclusion by P. Ehrenfest, it is a fact that Einstein missed its implications completely. First, he believed for a while that he had discovered a mistake in Friedmann's calculations, what he had to retract later, at the instance of the last, when he replied to Einstein that his calculations were perfectly right, and that it was Einstein who was in error. Had the creator of General Relativity, or maybe also W. de Sitter (who had himself already discovered an expanding solution, as early as 1917, although for a massless universe) realized these implications, they could have predicted the expansion of the universe from purely theoretical grounds, before astronomical evidence was there. Belenkiy sustains the opinion that, had this been the case, Einstein himself together with Friedmann and Slipher could had been solid candidates for a Nobel Prize in Physics  \cite{bel1}.

Lemaître could have qualified  for the Nobel Prize on his own since he played the role Einstein missed, namelyto connect Friedmann's theory, Slipher's redshifts and Hubble distances. As is well known, Lemaître had visited Hubble at Mount Wilson and Slipher at the Lowell Observatory, and graciously obtained the corresponding tables from each of them. Actually Lemaître (Hubble too) was indeed nominated for the  Nobel Prize in Physics in 1954 for ``his 1927 theoretical prediction of the expanding universe which was subsequently confirmed by the work of Hubble and Humason in the U.S.A.'' \cite{kra1}.
And for the expansion rate of the Universe, now called Hubble's constant, he had obtained a value very close to that of Hubble in the 1929 paper \cite{hubb1} (no wonder,  since both used practically the same data tables).

Finally, in January 1930, Eddington and de Sitter publicly rejected the very inspiration for Hubble's investigation `de Sitter's theory' as inadequate to explain the linear law! \cite{ds30} This, together with the fact that Lemaître had sent  Eddington again his 1927 paper, led soon the latter to re-consider at last Lemaître's work as a great discovery. An additional point, as explained in \cite{bel1}, is that both Hubble's and Lemaître's findings were made in spite of mistaken assumptions, and it happened that neither Lemaître immediately nor Hubble ever renounced their mistaken beliefs.
In a 1930 letter to Eddington, Lemaître writes (still not realizing the mistake in putting the spurious logarithmic term in his model, to make it unbounded in time):
``I consider a Universe of constant curvature in space but increasing with time and I emphasize the existence of a solution in which the motion of the nebulae is always a receding one from time minus infinity to plus infinity.'' \cite{nubi1} Later, Lemaître  dropped this term when Eddington pointed out that ``such logarithmic singularities have no physical significance'', and then started to consider the model with initial singularity, a prototype of the Big Bang model (the `monotone world of the first kind', in Friedmann's terminology). \cite{lem31,bel1}

As is now well known, Hubble never accepted the interpretation of his discovery as `expansion of the universe.' He tried to provide alternative explanations: \begin{quotation} {\it 
``it is difficult to believe that the velocities are real; that all matter is actually scattering away from our region of space. It is easier to suppose that the light waves are lengthened and the lines of the spectra are shifted to the red, as though the objects were receding, by some property of space or by forces acting on the light during its journey to the Earth.'' } \cite{hub29b,nubi1}
\end{quotation}
It is quite understandable that Hubble, together with the most substantial part of the astronomical community, were skeptical towards the idea of an expanding universe. This concept, which we nowadays consider almost trivial, was extremely difficult to accept at that time,  by astronomers and also  by theoreticians, even if Friedmann's and Lemaître's solutions were at hand. Indeed, in Hubble's case, though he recognized that the `expanding universe' could be a possibility, on which feasibility theoreticians, as de Sitter for one, had to decide (this is what he wrote in a letter to W. de Sitter), he considered that \begin{quotation} {\it  ``the 200-inch telescope will definitely answer the question of the interpretation of red-shifts, whether or not they represent actual motions, and if they do represent motions `if the universe
is expanding' the 200-inch may indicate the particular type of expansion.'' } \cite{hub_b1}
\end{quotation}

\section{Discussion and conclusion}

In this paper we have presented an original approach to the history of the formulation of the expansion law of the Universe, both in terms of perspective, which starts from the concept itself of what a physical theory is (this goes back to Galileo, as stressed in the introduction, but seems to be forgotten too easily in recent developments of theoretical physics), and new facts, which are put here together in such precise context for the first time. In this sense, the value of transmitting all this information, in a contextualized way, to non-specialists in the field should be appreciated. In particular, the strong connection of the pure mathematical equations, their symmetries and solutions, with their physical meaning, the usefulness of these solutions when they are confronted with precise and accurate experimental results, showing on its turn the same symmetry or pattern. The case at hand, the   expansion of the universe (by the way, one of the most important discoveries in the history of mankind) is maybe the most paradigmatic example one can find of the intimate interplay between theoretical model and experimental result in the construction of a physical theory. This has been made very clear in this paper. 

On one hand, Hubble's law was a purely empirical one and, on the other, Einstein was not clever enough as to appreciate the potentially enormous importance of Friedmann's  solution to his equations, which provided the theoretical counterpart. In his obsession, he even pretended to have found a mistake in  Friedmann's  solution! Finally, he realized he had missed the opportunity to predict the expansion of the universe on theoretical grounds. Had he been just a bit more clever than he was already, had he believed deeply in his equations as the only ones that could describe the Universe, he could have predicted himself that the Universe had to expand; since an expanding universe was a possible solution to its original equations (without $\Lambda$). Einstein confessed later that introducing $\Lambda$ was the worst mistake in his whole life (``die grösste Eselei meines Lebens''). No wonder, avoiding this too obvious, but also weird, move, he could have made an extraordinary prediction, even more important than those Wolfgang Pauli and Paul Dirac did, when they predicted, respectively, both based on purely theoretical considerations, the existence of the neutrino and the positron, long before they were discovered in the laboratory.

The discovery of expanding solutions to Einstein's GR equations was first made by W. de Sitter in 1917 (for the vacuum case, with cosmological constant) \cite{ds17}, and by A. Friedmann  in 1922 \cite{fri22}, and later by G. Lemaître in 1925 \cite{lem25}, for a more general expanding universe. Observational support for all these solutions was provided by the accurate redshift measurements of V.M. Slipher, starting  in 1912, and also by the much less precise (because of the inherent difficulties) distance measurements of E. Hubble in the 1920s. In 1927, Lemaître was the first to connect velocity and distance of spiral nebulae (the two tables, of Slipher and Hubble, respectively), and obtain what is now known as the Hubble constant, relating it with the expansion rate of his non-static solution of Einstein's equations---which he ignored had been previously found by Friedmann. The linear relationship between velocity and distance was confirmed by Hubble first in early 1929 \cite{hubb1} and later, in association with M. Humason, in 1931 \cite{huhu1}. Hubble's confirmation was not related with non-static solutions for the universe until early in 1930 when A. Eddington and W. de Sitter became fully aware of Lemaître's 1927 paper. As rightfully pointed out by Belenkiy \cite{bel1}: ``Errors can still lead to progress. The errors made on the way to the discovery of the expanding universe by the pioneers of modern cosmology did not prevent the cosmological community from getting finally the right physical interpretation.'' 

Furthermore, the first, key observation that led to conclude that the Universe is expanding was the fact that most of the spectra obtained by Slipher, over the years subsequent to his first measurement of the spectrum of the Andromeda Nebula (M31), showed a redshift, indicating velocity away from the observer. Even without distance measurements this could have led to an interpretation in terms of cosmic expansion. Slipher was too an extremely cautious and serious scientist, dedicated to scientific accuracy, to adventure possible interpretations without more definite proofs, and without excluding other possibilities. One of those was, e.g., that as all redshifts were obtained, at first, for objects visible from the northern hemisphere, it could well be that the results just indicated a movement of our solar system or the galaxy itself in the opposite direction. In the 1930s the importance of Slipher's results (and those of other astronomers, as well) were obscured by Hubble, who definitely liked to promote himself and the Mount Wilson Observatory, where he worked, as the first and only reference, when talking about the expansion law (he even dared to seriously advert W. de Sitter on that, in a letter written in 1930).

Having examined the most relevant literature, old and new, on the universe expansion issue, we conclude that a large share of the credit for the discovery of the expanding universe is due to Slipher, and yet, I fully coincide with Peacock that ``he tends to take very much second place to Hubble in most accounts.'' \cite{jpea1}  Slipher's other achievements in astronomy are also important, as his discovery of the atmospheric conditions of Mars, and his participation in the discovery of Pluto. \cite{Bremond2008}  
The cosmology community can debate whether reference to Lemaître should now be added to that to Hubble for the velocity-distance relation, but it seems fair to prominently acknowledge Slipher's pioneering contribution, as well. 
\medskip 

To conclude, going back to the beginning of this paper and trying to summarize as much as possible, my main points in vindicating Slipher, in the context of the IAU Resolution, are the following.
\ben
\item	In all the discussion, the IAU text of the resolution, and follow ups (as \cite{kra2}, for one) a very important issue has been systematically neglected: the very crucial role of the astronomer Vesto Slipher in the derivation of Hubble's law. To obtain the law, you need to compare two tables: one of radial velocities and one of distances to the extragalactic objects.
\item	While there is no doubt that Hubble produced the table of distances (he was a master in this respect, systematically using Henrietta Leavitt's law), the table of velocities was due to Vesto Slipher.
\item	Hubble himself was the first to recognize (albeit too late) the very important role of Slipher, talking of {\it ``your velocities and my distances''}, in a Letter of E.P. Hubble to V.M. Slipher, Mar 6, 1953 [Biographical Memoirs, Vol 52, National Academy of Sciences (U.S.)], and writing in his famous book E.P. Hubble, {\it ``The realm of the nebulae''} \cite{hub_b1}, while referring to Slipher that {\it ``... the first steps in a new field are the most difficult and the most significant. Once the barrier is forced further development is relatively simple.''} This is an explicit recognition of the seminal role of Slipher in the conception of the expansion law.
\item	Why did Hubble eventually say that? Although Slipher did not calculate the distances to the objects, his results on the redshifts, which he started to obtain in 1912, where so astonishing, pointing in some cases to such enormously high recession speeds, that he was undoubtedly the first astronomer to recognize that something weird, incredibly unusual was happening in the Universe: how could it be static at all if those objects were around, receding at such speeds? When he presented his results in the 1914 meeting of the American Astronomical Society chronicles say that he received a long, standing ovation! The community of astronomers (Hubble included) realized immediately that an important discovery was on the making.
\item	It may seem contradictory that in Lemaître's 1927 paper (now being vindicated) there is no mention to the table of redshifts by Vesto Slipher (what, by the way, adds reasons to his having been neglected!). This may seem very strange, since Lemaître perfectly knew about Slipher's results, from his visit to the Lowell observatory in Arizona during the period of his MIT Thesis. Instead, in deriving Hubble's law Lemaître takes his radial velocities from a table due to G. Strömberg \cite{stro1}. But, alas, you need only read the first page, even just the two first lines of Strömberg's paper to realize that all of it is, again, an extraordinary tribute to Slipher: {\it ``the great majority of the determinations being by Slipher.''} Obtained {\it ``through the perseverance of Professor M. Slipher.''} And so on.
\een

Summing up, 
\ben
\item	I consider the role of Slipher in the derivation of Hubble's law to be of paramount importance, as recognized (implicitly and explicitly), to begin with, by the two other actors of this drama, and subsequently by an increasing number of reputed specialists. His role was invaluable, both in inspiring the whole development (Hubble's dixit) and in providing one of the two tables that are absolutely necessary for the formulation of the law, both by Hubble and Lemaître (what nobody can oppose).
\item I therefore propose to the community of cosmologists that we re-name the Hubble law as Hubble-Lemaître-Slipher (HLS) law. In our naming of the law of the Universe expansion we can improve the IAU  original idea and give due credit to the three main actors of this play. 
\item	I am absolutely sure that both Edwin Hubble and Georges Lemaître would had been extremely happy with this decision.
\item	Further, I am also sure that, if the three brilliant cosmologists were alive {\it now}, under the standard criteria of the Nobel Academy, they would be perfect candidates for a shared Nobel Prize for their work that led to the discovery of the Universe expansion law.
\een

\medskip

\noindent {\bf Acknowledgement}. The author was partially supported by MINECO (Spain), FIS2016-76363-P,  by the CPAN Consolider Ingenio 2010 Project, and by AGAUR (Catalan Government), project 2017-SGR-247.

\end{document}